\documentclass[aps,prl,10pt,twocolumn,superscriptaddress]{revtex4-1}

\usepackage{amsmath}
\usepackage{graphicx}
\usepackage{amssymb}
\usepackage{hyperref}
\usepackage{times}
\usepackage{color}
\usepackage{bbold}

\hypersetup{
colorlinks=true,
linkcolor=blue,          
citecolor=blue,          
filecolor=magenta,       
urlcolor=cyan
}

\begin{document}

\title{Quantum-Limited Amplification and Parametric Instability in the Reversed Dissipation Regime of Cavity Optomechanics}

\author{A.~Nunnenkamp}
\affiliation{Department of Physics, University of Basel, Klingelbergstrasse 82, CH-4056 Basel, Switzerland}
\author{V.~Sudhir}
\affiliation{\'Ecole Polytechnique F\'ed\'erale de Lausanne (EPFL), CH-1015 Lausanne, Switzerland}
\author{A.~K.~Feofanov}
\affiliation{\'Ecole Polytechnique F\'ed\'erale de Lausanne (EPFL), CH-1015 Lausanne, Switzerland}
\author{A.~Roulet}
\affiliation{\'Ecole Polytechnique F\'ed\'erale de Lausanne (EPFL), CH-1015 Lausanne, Switzerland}
\author{T.~J.~Kippenberg}
\affiliation{\'Ecole Polytechnique F\'ed\'erale de Lausanne (EPFL), CH-1015 Lausanne, Switzerland}

\date{\today}

\begin{abstract}
Cavity optomechanical phenomena, such as cooling, amplification or optomechanically induced transparency, emerge due to a strong imbalance in the dissipation rates of the parametrically coupled electromagnetic and mechanical resonators. Here we analyze the reversed dissipation regime where the mechanical energy relaxation rate exceeds the energy decay rate of the electromagnetic cavity. We demonstrate that this regime allows for mechanically-induced amplification (or cooling) of the electromagnetic mode. Gain, bandwidth, and added noise of this electromagnetic amplifier are derived and compared to amplification in the normal dissipation regime. In addition, we analyze the parametric instability, i.e.~optomechanical Brillouin lasing, and contrast it to conventional optomechanical phonon lasing. Finally, we propose an experimental scheme that realizes the reversed dissipation regime using parametric coupling and optomechanical cooling with a second electromagnetic mode enabling quantum-limited amplification. Recent advances in high-$Q$ superconducting microwave resonators make the reversed dissipation regime experimentally realizable.
\end{abstract}


\maketitle

\emph{Introduction.} The parametric coupling between electromagnetic (e.g.~optical or microwave) and mechanical resonators -- first studied in the context of gravitational-wave detection \cite{Braginsky1977} -- forms the basis of a variety of cavity optomechanical phenomena, which have been intensely studied in recent years \cite{Aspelmeyer2013, Kippenberg2008}. For example, it has led to readout of mechanical motion with an imprecision below that at the standard quantum limit \cite{Teufel2009, Anetsberger2010, Westphal2012}. The dynamical backaction associated with this coupling is the basis for optomechanical sideband cooling \cite{Wilson-Rae2007, Marquardt2007, Schliesser2008} that has proven to be the enabling route to cool a mechanical oscillator to its quantum ground-state and realize low-entropy mechanical oscillator states \cite{Teufel2011, Chen2011, Verhagen2012}. Driving on the upper mechanical sideband enables amplification of mechanical motion, which can lead to phonon lasing, i.e.~self-sustained mechanical oscillations \cite{Kippenberg2005, Marquardt2006, Vahala2009}, and it has been used to amplify microwave signals \cite{Massel2011}. Moreover, optomechanical coupling is the basis of optomechanically-induced transparency \cite{Weis2010, 2011_Chen_SinglePhotonDevices, Zhang2003}, an effect that can be used in both the classical and quantum regime to slow, advance, and store electromagnetic fields in mechanical degrees of freedom \cite{2011_Painter_EITOptomechanics, Zhou2013, Palomaki2013}.

\begin{figure}
\includegraphics[width=0.9\columnwidth]{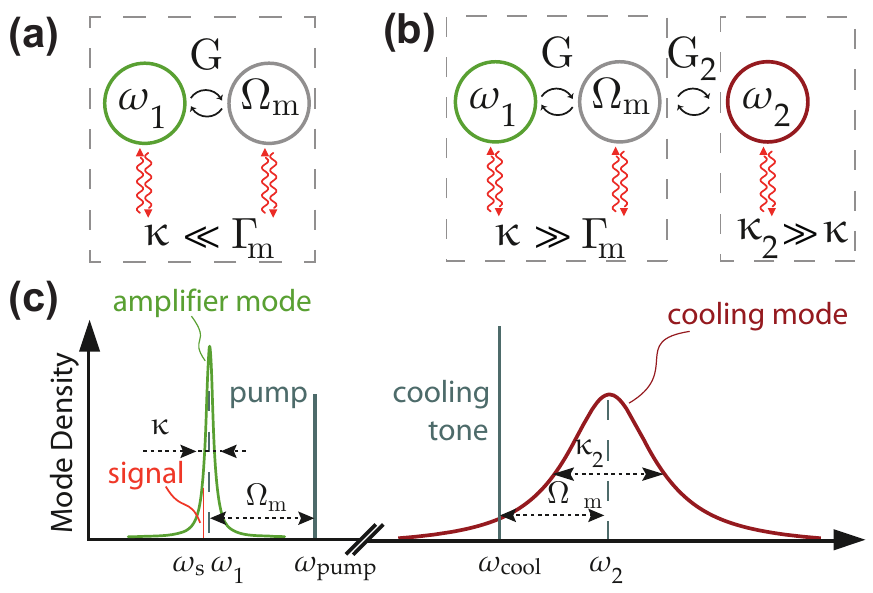}
\caption{(Color online) \emph{Reversed dissipation regime (RDR).} (a) For a single electromagnetic mode coupled to a mechanical oscillator this hierarchy requires the mechanical damping rate to greatly exceed the electromagnetic damping rate ($\Gamma_{m} \gg \kappa$). (b) Realization of the RDR using two electromagnetic modes with strongly different energy decay rates ($\kappa_{2} \gg \kappa$). In this scenario the RDR can be reached by optomechanical sideband cooling a high-$Q$ mechanical oscillator with the dissipative electromagnetic mode $\kappa_{2}$. This can establish the RDR with respect to the low dissipation electromagnetic mode $\kappa$. (c) Experimental scheme to realize optomechanical amplification in the RDR with two electromagnetic modes with different damping rates. The signal frequency $\omega_S$ is close to the amplifier mode frequency $\omega_1$.}
\label{fig:setup}
\end{figure}

All the cavity optomechanical phenomena mentioned above rely on a hierarchy of time scales in which the mechanical degree of freedom is damped at a rate that is much smaller than that of the electromagnetic cavity. This is also generally the case in all experimentally realized optomechanical systems to date. In this paper we introduce the notion of a \emph{reversed dissipation regime} of cavity optomechanics, where the mechanical dissipation rate ($\Gamma_{m}$) is much larger than the cavity linewidth ($\Gamma_{m}\gg\kappa$). While this does not enable us to use the mechanical oscillator as a high-coherence element, the mechanical oscillator will play the helpful role of an additional dissipative reservoir for the electromagnetic mode \cite{Wang2013, Metelmann2013}.

To be precise, we demonstrate that in the reversed dissipation regime it is the linewidth of the electromagnetic mode that is strongly affected by the optomechanical interaction. If the cavity is driven on the upper mechanical sideband, the system becomes a nearly ideal phase-preserving amplifier of electromagnetic input signals. It has a large gain-bandwidth product given by the cavity linewidth and can reach the quantum limit, i.e.~it amplifies electromagnetic signals in the limit of large gain while adding only half a quantum of noise \cite{Caves1982a}. Subsequently, we analyze the self-sustained oscillations above the instability threshold, that are dominated by the electromagnetic mode. In this Brillouin maser \cite{Li2013, Bahl2011}, mechanical sidebands in the electromagnetic output field are strongly suppressed and the oscillation frequency is shifted from the pump frequency by an amount that deviates from the mechanical frequency. We propose an implementation in an electromechanical setup with superconducting circuits \cite{You2011, Devoret2013}, where two electromagnetic modes (with largely different decay rates) are coupled to one mechanical oscillator. It has the advantage of reaching the reversed dissipation regime while simultaneously quenching thermal fluctuations in the mechanical oscillator. Our study opens a practical way to achieve a quantum-limited optomechanical amplifier with a large gain-bandwidth product complementing phase-sensitive amplifiers based on superconducting Josephson junctions that are operating at the quantum limit \cite{Yurke1988, Castellanos-Beltran2008, Bergeal2010}.

\emph{Model.} We consider the standard optomechanical system, i.e.~a mechanical oscillator whose position modulates the frequency of an electromagnetic mode. The classical nonlinear equations of motion in the frame of the cavity drive are \cite{SI}
\begin{eqnarray}
\dot{\bar{a}} & = & +i\Delta_{0}\bar{a}-\frac{\kappa}{2}\bar{a}+ig_{0}\bar{a}(\bar{b}+\bar{b}^{\star}) +\sqrt{\kappa} \, \bar{a}_\textrm{in} \label{eq:classicala}\\
\dot{\bar{b}} & = & -i\Omega_{m}\bar{b}-\frac{\Gamma_{m}}{2}\bar{b}+ig_{0}|\bar{a}|^{2}
\label{eq:classicalb}
\end{eqnarray}
where $-i \Omega = \sqrt{\kappa} \, \bar{a}_\textrm{in}$, $\bar{a}$ and $\bar{b}$ are the cavity and mechanical amplitudes, $\Delta_0$ the detuning, $\kappa$ and $\Gamma_m$ the cavity and mechanical dissipation rates, $g_0$ the single-photon optomechanical coupling, $\Omega_m$ the mechanical frequency, and $\Omega$ the drive strength. Fluctuations around the amplitudes $\bar{a}$ and $\bar{b}$ are described by bosonic operators $\delta \hat{a}$ and $\delta \hat{b}$ obeying linear quantum Langevin equations \cite{SI}
\begin{eqnarray}
\dot{\delta\hat{a}} & = & +i\Delta \delta\hat{a} -\frac{\kappa}{2} \delta\hat{a} +iG(\delta\hat{b}+\delta\hat{b}^{\dagger})+\sqrt{\kappa}\,\hat{a}_{\text{in}} \label{eq:langevind}\\
\dot{\delta\hat{b}} & = & -i\Omega_{m} \delta\hat{b}-\frac{\Gamma_{m}}{2}\delta\hat{b} +iG(\delta\hat{a}+\delta\hat{a}^{\dagger})+\sqrt{\Gamma_{m}}\,\hat{b}_{\text{in}} \label{eq:langevinc}
\end{eqnarray}
where $\Delta=\Delta_{0}+g_{0}(\bar{b}+\bar{b}^{\star})$ denotes the shifted detuning and $G=g_{0}\bar{a}$ is the enhanced optomechanical coupling.
$\hat{a}_{\textrm{in}}$ and $\hat{b}_{\text{in}}$ are zero-mean input noise operators with $[\hat{a}_{\textrm{in}}(t), \hat{a}_{\text{in}}^\dagger(t')] = [\hat{b}_{\textrm{in}}(t), \hat{b}_{\textrm{in}}^\dagger(t')] = \delta(t-t')$, $\langle \hat{a}_{\text{in}}(t) \hat{a}_{\text{in}}^\dagger(t')\rangle = \delta(t-t')$, and $\langle \hat{b}_{\text{in}}^\dagger(t) \hat{b}_{\text{in}}(t')\rangle = \bar{n}_\textrm{th} \delta(t-t')$ and $\bar{n}_\textrm{th}$ is the thermal phonon number. We only consider the external coupling to the electromagnetic feedline, but including additional loss channels is straightforward. Our treatment of mechanical dissipation is correct for large quality factors, i.e.~$\Gamma_{m}\ll\Omega_{m}$. In the reversed dissipation regime, $\kappa\ll\Gamma_{m}$, this implies the strongly resolved sideband limit, i.e. $\kappa \ll \Gamma_{m} \ll \Omega_{m}$. We first focus on an implementation with one electromagnetic mode, cf.~Fig.~\ref{fig:setup} (a), and discuss an implementation with two modes, cf.~Fig.~\ref{fig:setup} (b) and (c), below.

We see that the equations of motion for the fluctuations (\ref{eq:langevind}) and (\ref{eq:langevinc}) are, up to variances of the noise inputs, symmetric with respect to interchanging the electromagnetic and mechanical degrees of freedom. As a consequence, optomechanical phenomena in the normal dissipation regime (NDR) have corresponding partners in the reversed dissipation regime (RDR). For example, cooling and amplification of the mechanical oscillator by the cavity field in the NDR translate to cooling and amplification of the optical or microwave degree of freedom by the mechanics in the RDR. However, the classical nonlinear equations of motion (\ref{eq:classicala}) and (\ref{eq:classicalb}) do not respect this symmetry. This will have important consequences for the nonlinear dynamics in the lasing regime that we will discuss below.

\begin{figure}
\includegraphics[width=\columnwidth]{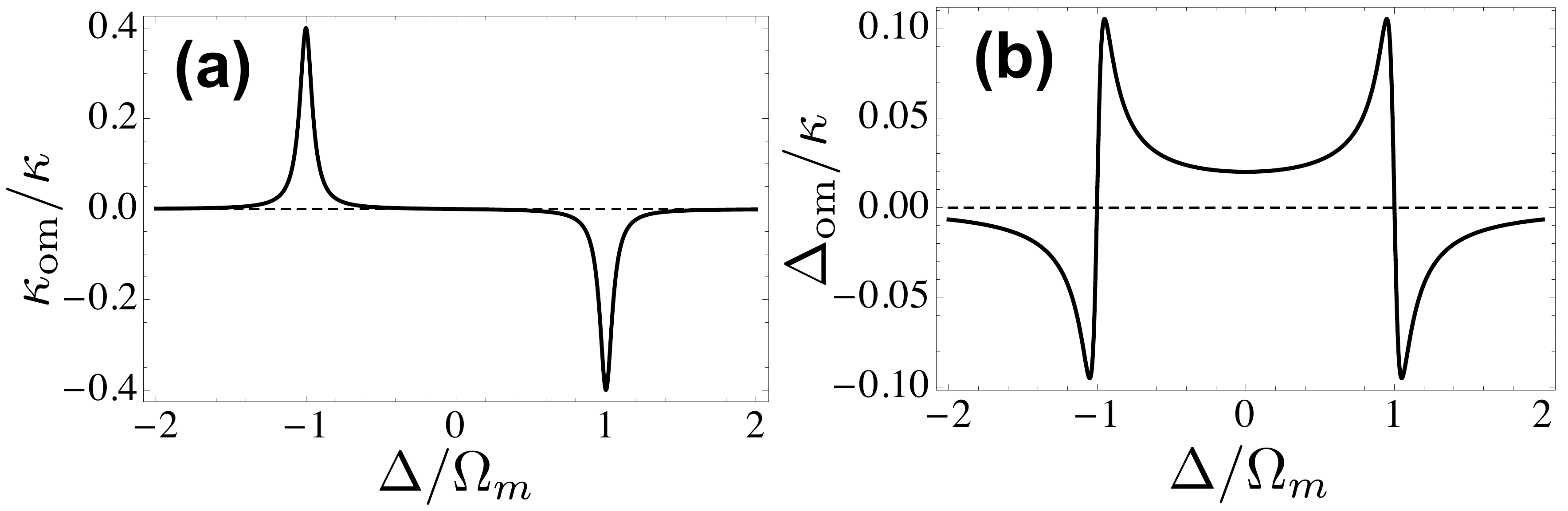}
\caption{\emph{Dynamical backaction effects on the electromagnetic mode in the RDR.}
(a) Optomechanically-induced shift of the cavity linewidth $\kappa_{\textrm{om}}/\kappa$
(\ref{eq:kappaeff}) and (b) the cavity frequency $\Delta_{\textrm{om}}/\kappa$
as a function of detuning $\Delta$. The parameters are $\Omega_{m}/\kappa=10^{4}$
and $\Omega_{m}/\Gamma_{m}=G/\kappa=10$.}
\label{fig:kappaeff} 
\end{figure}

\emph{Quantum noise approach.} For weak coupling and in the RDR, $\Gamma_m \gg \kappa$, where the mechanical oscillator is much faster damped than the cavity mode, we are able to treat the mechanical oscillator as a quantum noise source for the cavity. The optomechanical interaction will induce transitions from the states with $n$ photons to the states with $n\pm1$ photons. Utilizing Fermi's Golden Rule, to second order in the coupling $G$, these rates are $\Gamma_{n\rightarrow n+1}=(n+1)\Gamma_{\uparrow}$ and $\Gamma_{n\rightarrow n-1}=n\Gamma_{\downarrow}$ with $\Gamma_{\uparrow}=G^{2}S_{xx}(-\Delta)/x_{\textrm{ZPF}}^{2}$ and $\Gamma_{\downarrow}=G^{2}S_{xx}(+\Delta)/x_{\textrm{ZPF}}^{2}$ where $\hat{x}=x_{\textrm{ZPF}}(\delta\hat{b}+\delta\hat{b}^{\dagger})$ is the displacement of the mechanical oscillator, and $S_{xx}(\omega)$ is its spectral density. The effective cavity linewidth is $\kappa_{\textrm{eff}}=\kappa+\kappa_{\textrm{om}}$ with $\kappa_{\textrm{om}}=\Gamma_{\downarrow}-\Gamma_{\uparrow}$ given by 
\begin{eqnarray}
\kappa_{\textrm{om}}=\frac{\Gamma_{m}G^{2}}{\frac{\Gamma_{m}^{2}}{4}+(\Delta+\Omega_{m})^{2}}-\frac{\Gamma_{m}G^{2}}{\frac{\Gamma_{m}^{2}}{4}+(\Delta-\Omega_{m})^{2}}.\label{eq:kappaeff}
\end{eqnarray}
Analogous to the NDR we have $\kappa_{\textrm{om}}=-2\,\textrm{Im}\Sigma[-\Delta]$ where the self energy is $\Sigma[\omega]=-iG^{2}(\chi_{m}[\omega]-\chi_{m}[-\omega]^{\star})$ and the mechanical response $\chi_{m}[\omega]=[\Gamma_{m}/2-i(\omega-\Omega_{m})]^{-1}$ \cite{SI}. The optomechanically-induced damping is accompanied by a frequency shift of the cavity mode (a ``mechanical spring'' effect) so that $\Delta_{\textrm{eff}}=\Delta+\Delta_{\textrm{om}}$ with $\Delta_{\textrm{om}}=-\textrm{Re}\,\Sigma[-\Delta]$.

In Fig.~\ref{fig:kappaeff} we plot the optomechanically-induced shift of the cavity linewidth $\kappa_{\textrm{om}}$ and the cavity frequency $\Delta_{\textrm{om}}$ as a function of the detuning $\Delta$. On the red sideband $\Delta=-\Omega_{m}$ the cavity linewidth is increased by the coupling to the mechanical oscillator which plays the role of an additional dissipative channel. Depending on the thermal phonon and photon numbers this can lead to cooling of the electromagnetic degree of freedom. In contrast, driving on the blue sideband $\Delta=+\Omega_{m}$ the cavity linewidth $\kappa_{\textrm{eff}}$ is reduced leading to optomechanical amplification, i.e.~here the strongly dissipative mechanical oscillator introduces anti-damping of the cavity mode. We also see that in the resolved sideband limit the shift in the cavity frequency $\Delta_{\textrm{om}}$ is small close to the mechanical sidebands.

\emph{Amplification: gain, bandwidth, and added noise.} To study the amplification and de-amplification properties we solve the Langevin equations (\ref{eq:langevind}) and (\ref{eq:langevinc}) together with the input-output relation, $\hat{a}_{\textrm{out}}=\hat{a}_{\textrm{in}}-\sqrt{\kappa} \, \delta\hat{a}$ , see e.g.~Ref.~\cite{Clerk2008a}. The result has the form \cite{SI}
\begin{equation}
\label{eq:inoutsolution}
\hat{a}_{\textrm{out}}=A(\omega)\hat{a}_{\textrm{in}}+B(\omega)\hat{a}_{\textrm{in}}^{\dagger}+C(\omega)\hat{b}_{\textrm{in}}+D(\omega)\hat{b}_{\textrm{in}}^{\dagger}.
\end{equation}

\begin{figure}
\includegraphics[width=\columnwidth]{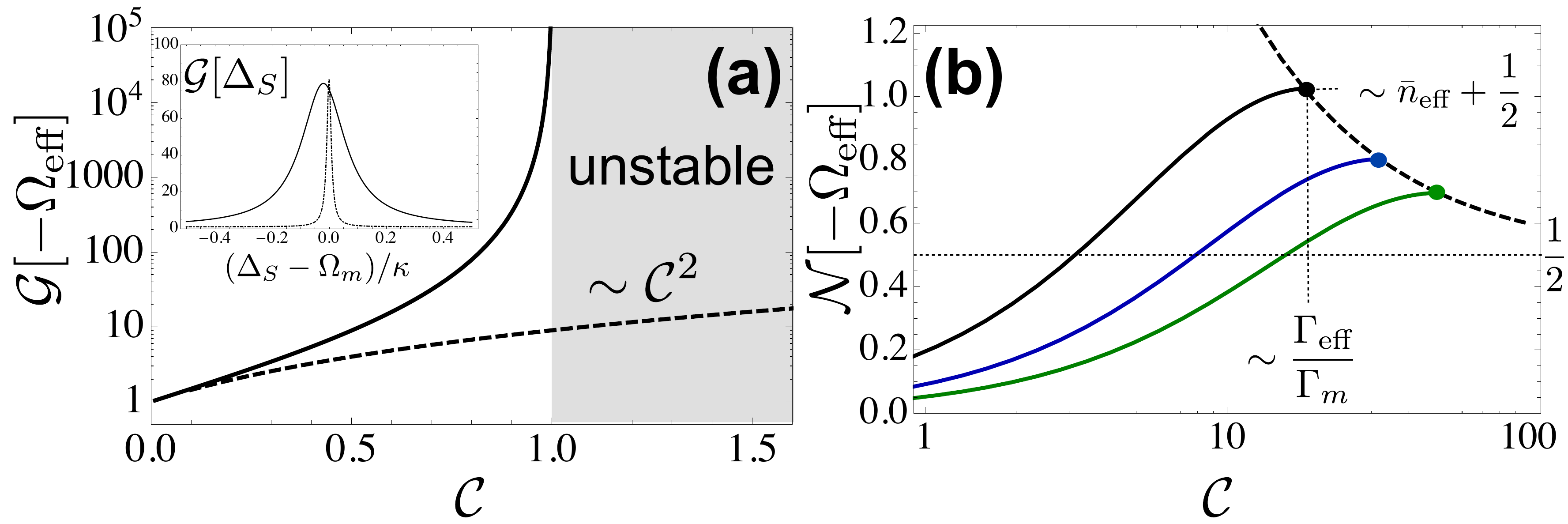}
\caption{(Color online) \emph{Quantum-limited amplification.}
(a) Gain on resonance $\mathcal{G}[-\Omega_{\textrm{eff}}]$ in the one-mode RDR as a function of the cooperativity $\mathcal{C}= 4G^{2} / (\Gamma_{m}\kappa)$ for $\Delta=\Omega_{m}$, $\Gamma_{m}/\kappa=10$, $\Omega_{m}/\kappa=50$ (solid). The regime of self-sustained oscillations is marked in grey. In the inset we plot the gain $\mathcal{G}[\Delta_{S}]$ as a function of signal frequency for $\mathcal{C}=0.8$ and $\Gamma_{m}/\kappa=10$ (RDR) (solid) as well as $\Gamma_{m}/\kappa=0.1$ (NDR) (dash-dotted). We stress the much larger amplification bandwidth in the RDR relative to the one in the NDR. In addition, we show the case of the two-mode RDR, cf.~Fig.~\ref{fig:setup} (b). For $G_2 \gg G$ this mode can induce a transition from the NDR to the RDR via optomechanical cooling. This will displace the point of parametric instability to $\mathcal{C} = \Gamma_\textrm{eff}/\Gamma_m$. The case $\kappa = \kappa_{2}$ and $G=G_{2}$ (dashed) leads to a cancellation of the dynamical backaction ($\Gamma_\textrm{eff} = \Gamma_m$) and to an amplification mechanism with unlimited gain-bandwidth product and without an instability threshold \cite{Metelmann2013}. (b) Added noise on resonance $\mathcal{N}[-\Omega_{\textrm{eff}}]$ as a function of cooperativity $\mathcal{C}$ for the two-mode RDR with $\bar{n}_{\textrm{th}} = 10$ and $\kappa_2 = 5\kappa$, (solid, from left to right) for $G_{2}/\kappa=0.1,0.2,0.3$ as well as $\kappa_{2} = \kappa$ and $G_{2}=G$ (dashed).}
\label{fig:amplifier}
\end{figure}

In the resolved sideband limit $\Omega_{m}\gg\kappa$, we have $|B(\omega)|\ll|A(\omega)|$ and $|C(\omega)|\ll|D(\omega)|$, i.e.~the amplifier is (close to) phase preserving (phase insensitive) \cite{Caves1982a, Clerk2008a}. It is thus characterized by the power gain $\mathcal{G}=|A|^{2}$ and the number of added noise photons $\mathcal{N}=(\bar{n}_{\textrm{eff}}+\frac{1}{2})\frac{|D|^{2}}{|A|^{2}}$ at the signal frequency $\omega_{S}$ in the frame of the pump $\omega_{P}$, i.e.~$\Delta_{S}=\omega_{S}-\omega_{P}$, where $\bar{n}_{\textrm{eff}}$ is the effective phonon number of the mechanical oscillator.

For weak coupling the gain $\mathcal{G}$ for a signal at $\Delta_{S}$ is determined by the effective linewidth $\kappa_{\textrm{eff}}$ and frequency shift $\Delta_{\textrm{eff}}$
\begin{equation}
\mathcal{G}[\Delta_{S}]=\left|1-\frac{\kappa}{\frac{\kappa_{\textrm{eff}}}{2}-i(\Delta_{S}+\Delta_{\textrm{eff}})}\right|^{2}.
\end{equation}

In the resolved sideband limit the counterrotating terms in Eqs.~(\ref{eq:langevind}) and (\ref{eq:langevinc}) can be neglected and become those of a non-degenerate parametric amplifier, leading to $\kappa_{\textrm{om}}=4G^{2}/\Gamma_{m}$ and $\Delta_{\textrm{om}}=0$. In this case we obtain for the gain $\mathcal{G}=|(\mathcal{C}+1)/(\mathcal{C}-1)|^2$ which diverges as the system approaches the instability to parametric oscillations at a cooperativity $\mathcal{C}= 4G^{2}/(\Gamma_{m}\kappa)\rightarrow 1$. Importantly, the bandwidth of this amplifier is given by $\kappa_{\textrm{eff}}$ and its gain-bandwidth product is $\kappa$. The number of added noise photons of the amplifier on the shifted resonance, i.e.~$\Delta_{S} = -\Omega_\textrm{eff} = -\Omega_m + \textrm{Re}\,\Sigma[-\Delta]$, is
\begin{equation}
\mathcal{N}[-\Omega_\textrm{eff}]=\frac{4\mathcal{C}(\bar{n}_{\textrm{eff}}+\frac{1}{2})}{(\mathcal{C}+1)^{2}}\rightarrow\bar{n}_{\textrm{eff}}+\frac{1}{2}
\end{equation}
for $\mathcal{C}\rightarrow$ 1, i.e.~the amplifier reaches the quantum limit for a phase-preserving amplifier close to instability where $\kappa_{\textrm{eff}}$ goes to zero if the (effective) thermal phonon number is zero $\bar{n}_{\textrm{eff}}=0$ and if there are no additional loss channels.

In Fig.~\ref{fig:amplifier} (a) we plot the gain $\mathcal{G}[-\Omega_\textrm{eff}]$ for the pump on the blue mechanical sideband $\Delta=\Omega_{m}$ and the signal at the cavity resonance featuring large gain close to the parametric instability to self-sustained oscillations. The maximum gain is limited by the sideband parameter $\Omega_{m}/\kappa$, but can be increased by shifting the signal frequency slightly away from resonance.

These properties should be compared with optomechanical amplification in the NDR, $\kappa \gg \Gamma_m$, as shown in Ref.~\cite{Massel2011}. In the resolved-sideband limit that system maps on a non-degenerate parametric amplifier, i.e.~the gain is identical to the one in the RDR shown in Fig.~\ref{fig:amplifier} (a). A striking difference between these two amplifiers, however, is their gain-bandwidth product. As seen from the inset of Fig.~\ref{fig:amplifier} (a), in the NDR it is given by the mechanical linewidth $\Gamma_{m}$ which is typically orders of magnitude smaller than the cavity linewidth $\kappa$.

It is advantageous to realize the RDR not with a low-$Q$ mechanical oscillator that consequentially has a large mechanical decoherence rate $\Gamma_{m}\bar{n}_{\textrm{th}}$, but with a high-$Q$ mechanical oscillator whose damping is increased by sideband cooling \cite{Schliesser2008}, i.e.~a mechanical oscillator with large quality factor $\Gamma_{m}\ll\kappa$ is coupled to a second electromagnetic mode whose linewidth greatly exceeds linewidth of the first mode $\kappa_{2}\gg\kappa$ while satisfying the resolved-sideband condition $\Omega_{m}\gg\kappa_{2}$, cf.~Fig.~\ref{fig:setup} (b) and (c). For more details see Supplementary Material \cite{SI}.

Driving this mode on the lower mechanical sideband leads to cooling of the mechanical mode with damping rate $\Gamma_{\textrm{eff}}=(\mathcal{C}_{2}+1)\Gamma_{m}$. For $\mathcal{C}_{2}=4G_{2}^{2}/(\kappa_{2}\Gamma_{m})\gg1$, but still within the weak coupling limit, the effective mechanical linewidth is $\Gamma_{\textrm{eff}}\sim\kappa_{2}/2$, limited by the onset of parametric normal-mode splitting \cite{Dobrindt2008}. This cooling therefore establishes the RDR with respect to the first electromagnetic mode as the mechanical damping satisfies $\Gamma_{\textrm{eff}}\sim\kappa_{2}/2\gg\kappa$. (The regime of normal-mode splitting is not considered here.) In addition to reaching the RDR, this has the distinct advantage of cooling the mechanical oscillator to an effective phonon number (granted that $\kappa \gg \Gamma_m \bar{n}_m$) of
\begin{equation}
\bar{n}_{\textrm{eff}}=\frac{\bar{n}_{\textrm{th}}}{\mathcal{C}_{2}+1}+\frac{\mathcal{C}_2}{\mathcal{C}_2+1} \frac{\kappa_{2}^{2}}{16\Omega_{m}^{2}}.
\end{equation}

In Fig.~\ref{fig:amplifier} (b) we see that for $\mathcal{C} \rightarrow \Gamma_\textrm{eff}/\Gamma_m$, i.e.~close to the instability threshold, the amplifier approaches $\mathcal{N} \rightarrow \bar{n}_\textrm{eff} + \frac{1}{2}$ which for increasing cooling power $G_2$ reaches the quantum limit at high gain, i.e.~$\mathcal{N}\rightarrow\frac{1}{2}$ \cite{Caves1982a, Clerk2008a}, as $\bar{n}_\textrm{eff}\rightarrow 0$. Note that we have expressed the cooperativity in terms of the bare mechanical damping rate $\Gamma_m$, i.e.~in the absence of cooling.

In the case $\kappa = \kappa_{2}$ and $G=G_{2}$ the dynamical radiation-pressure back-action effects of the two electromagnetic modes cancel, and the system remains stable independent of the coupling strength, i.e.~there is no parametric instability. In the resolved-sideband limit this realizes a quantum-limited phase-preserving amplifier with an unlimited gain-bandwidth product \cite{Metelmann2013}. In Fig.~\ref{fig:amplifier} (a,b) we also show gain $\mathcal{G}$ and noise $\mathcal{N}$ in this case for comparison.

\begin{figure}
\includegraphics[width=0.8\columnwidth]{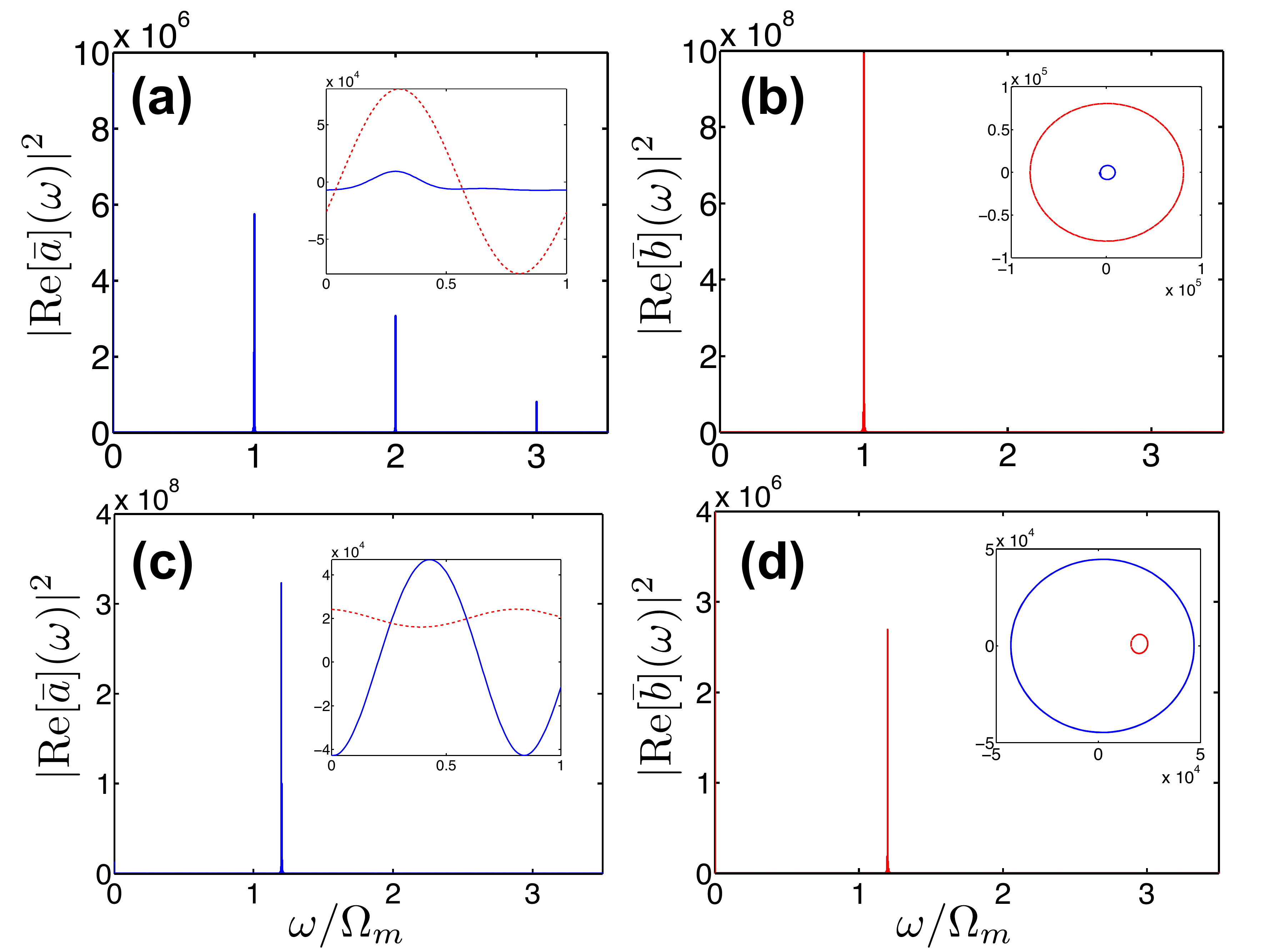}
\caption{(Color online) \emph{Self-sustained oscillations.} Electromagnetic $\textrm{Re}[\bar{a}]$ (blue) and mechanical mode $\textrm{Re}[\bar{b}]$ (red) in NDR ($\Gamma_{m}/\Omega_{m}=0.001$, $\kappa/\Omega_{m}=0.1$) in (a, b) and RDR ($\Gamma_{m}/\Omega_{m}=0.1$, $\kappa/\Omega_{m}=0.001$) in (c, d). The main plots show Fourier transforms $|\textrm{Re}[\bar{a}](\omega)|^{2}$ and $|\textrm{Re}[\bar{b}](\omega)|^{2}$. Insets in (a, c) show $\textrm{Re}[\bar{a}]$ (solid blue) and $\textrm{Re}[\bar{b}]$ (dashed red) as a function of time $\Omega_{m}t/(2\pi)$, insets in (b, d) $\textrm{Re}[\bar{a}]$ vs.~$\textrm{Im}[\bar{a}]$ (solid blue) and $\textrm{Re}[\bar{b}]$ vs.~$\textrm{Im}[\bar{b}]$ (dashed red). We set $\Omega/\Omega_{m}=2.5\times10^{3}$, $g_{0}/\Omega_{m}=10^{-5}$, and $\Delta/\Omega_{m}=0.8$.}
\label{fig:instability} 
\end{figure}

\emph{Parametric instability and nonlinear oscillations.} When the effective electromagnetic (mechanical) linewidth $\kappa_{\textrm{eff}}$ ($\Gamma_\textrm{eff}$) in the RDR (NDR) goes to zero, the system undergoes a lasing transition to self-sustained oscillations.
In Fig.~\ref{fig:instability} (a) we consider the NDR ($\kappa\gg\Gamma_{m}$) for one electromagnetic mode. Solving for the long-time dynamics of the nonlinear classical equations (\ref{eq:classicala}) and (\ref{eq:classicalb}) we see the mechanics oscillates at its natural frequency $\Omega_{m}$ and the modulus of the mechanical amplitude is much larger than the electromagnetic one. The cavity mode has several frequencies that are multiples of its mechanical frequency $\Omega_{m}$ \cite{Marquardt2006}.

Self-sustained oscillations in the RDR, $\kappa\ll\Gamma_{m}$, are shown in Fig.~\ref{fig:instability} (b). The mechanical and cavity amplitudes oscillate at a frequency $\Delta_{0}+g_{0}\Re[\bar{b}]$, the electromagnetic amplitude is much larger than the mechanical amplitude, and there is a single dominant frequency in the electromagnetic mode.

Therefore, the parametric instability in the RDR enables to achieve a spectrally pure electromagnetic tone similarly to a Brillouin Maser or Laser \cite{Li2013, Bahl2011}. The linewidth of the generated electromagnetic radiation will decrease inversely with the number of quanta in the laser mode and exhibit a fundamental linewidth contribution from quantum backaction \cite{VahalaPRA2008}. The self-sustained oscillations in the RDR can thus serve as high-coherence oscillators.

\emph{Experimental feasibility}. Recent advances in the fabrication of ultra high-$Q$ microwave cavities make the experimental realization of the two-mode RDR discussed above a realistic endeavor. In the microwave domain titanium nitride (TiN) coplanar waveguide cavities and lumped element resonators have been demonstrated with unloaded quality factors above $3\times10^{7}$ at large photon numbers \cite{Leduc2010aa, Vissers2010aa} which corresponds to cavity linewidths of $\kappa/2\pi\sim\mathcal{O}(100\,\mathrm{Hz})$. The lumped element microwave cavities can be strongly coupled to the mechanical drum modes of suspended vacuum-gap capacitors with resonance frequencies in the 10 MHz range \cite{Teufel2011} featuring vacuum optomechanical coupling rates of $g_{0}/2\pi\sim\mathcal{O}(100\,\mathrm{Hz})$.

\begin{figure}
\includegraphics[width=0.8\columnwidth]{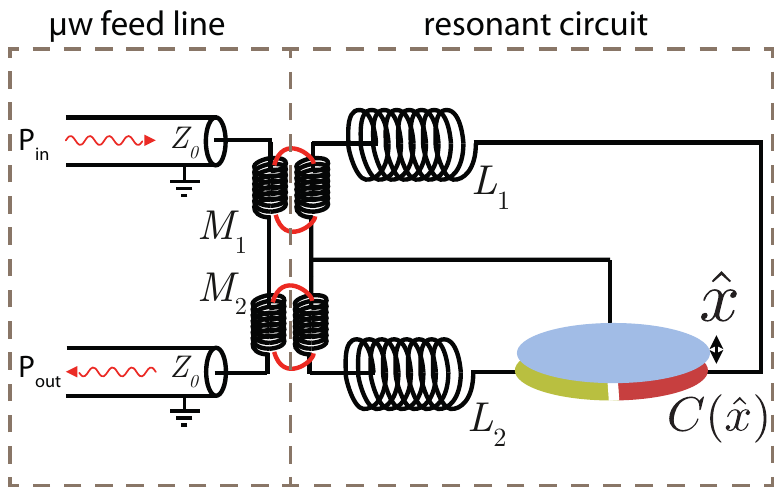}
\caption{(Color online) \emph{Suggested experimental scheme using superconducting circuits.} Coupling the mechanical element $\hat{x}$ realized as the moving plate of a capacitor $C(\hat{x})$ simultaneously to a high-$Q$ and a low-$Q$ microwave cavity. The bottom electrode of the capacitor is split into two equal parts that are connected to two different inductances and thus constitute two lumped element microwave resonators. Their very different quality factors arise from different strengths of inductive coupling to the feedline.}
\label{fig:circuit}
\end{figure}

To realize the two-mode RDR the mechanical oscillator is coupled \emph{simultaneously} to a high-$Q$ and a low-$Q$ microwave resonance. This can be achieved with the circuit seen in Fig.~\ref{fig:circuit}. It consists of two $LC$-resonators with different resonance frequencies and different mutual inductances between resonators and the feedline, i.e.~one mode can be strongly while the other weakly coupled $\kappa_{2}\gg\kappa$. Assuming $g_{0}/2\pi\sim\mathcal{O}(100\,\mathrm{Hz})$, $\kappa/2\pi=10\,\mathrm{kHz}$, and $\kappa_{2}/2\pi=200\,\mathrm{kHz}$, the critical mean photon number $\bar{n}_{p}$ to undergo self-sustained oscillations is $4g_{0}^{2}\bar{n}_{p}/(\kappa\Gamma_{\mathrm{eff}})>1$, i.e.~$\bar{n}_{p}\sim\mathcal{O}(10^{4})$, which superconducting microwave cavities can sustain. We note that in contrast to amplifiers based on Josephson junctions \cite{Bergeal2010}, our device does not require any external magnetic fields and is thus insensitive to magnetic field fluctuations.

\emph{Conclusion.} In this letter we introduced the notion of the reversed dissipation hierarchy where the mechanical dissipation rate is larger than the cavity linewidth. We showed how to use it as a quantum-limited phase preserving amplifier with large gain-bandwidth product. We also studied the self-sustained oscillations with strongly suppressed mechanical sidebands. Finally, we proposed an experimentally feasible circuit coupling two electromagnetic and one mechanical modes.

\emph{Acknowledgements.} We thank C.~Bruder for interesting discussions. This work was financially supported by the SNF and the NCCR Quantum Science and Technology. TJK acknowledges financial support from an ERC AdG (QREM), Marie Curie ITN cQOM, and the DARPA ORCHID program.

%


\begin{widetext}
\begin{center}
{\fontsize{12}{12}\selectfont
\textbf{Supplemental Material for ``Quantum-Limited Amplification and Parametric Instability in the Reversed Dissipation Regime of Cavity Optomechanics''\\[5mm]}}
{\normalsize A.~Nunnenkamp$^1$, V.~Sudhir$^2$, A.~K.~Feofanov$^2$, A.~Roulet$^2$, and T.~J.~Kippenberg$^2$\\[1mm]}
{\fontsize{9}{9}\selectfont  
\textit{$^1$Department of Physics, University of Basel, Klingelbergstrasse 82, CH-4056 Basel, Switzerland\\
$^2$\'Ecole Polytechnique F\'ed\'erale de Lausanne (EPFL), CH-1015 Lausanne, Switzerland}}
\end{center}
\normalsize
\end{widetext}

\section{Derivation of the equations of motion (1)-(4)}

In this section we will derive the classical nonlinear equations for motion for the mean field and the linearized EOM for the fluctuations, i.e.~Equations~(\ref{eq:classicala}) to (\ref{eq:langevinc}) of the main text.

We consider the standard optomechanical system, i.e.~a mechanical oscillator whose position modulates the frequency of an electromagnetic mode. The Hamiltonian ($\hbar = 1$) reads \cite{Aspelmeyer2013}
\begin{equation}
H_0 = \omega_R \hat{a}^\dagger \hat{a} + \Omega_m \hat{b}^\dagger \hat{b} - g_0 \hat{a}^\dagger \hat{a} (\hat{b}+\hat{b}^\dagger)
\end{equation}
where $\omega_R$ is the resonator frequency, $\Omega_m$ the mechanical frequency, $g_0$ the single-photon optomechanical coupling, and $\hat{a}$ and $\hat{b}$ satisfy canonical bosonic commutation relations.

Let us consider the open quantum system where the electromagnetic mode $\hat{a}$ is driven coherently by a laser or microwave source. This situation can be described by adding the following term to the Hamiltonian $H_d = \Omega (\hat{a}^\dagger e^{-i \omega_L t} + \textrm{H.c.})$ where $\Omega$ is the drive strength and $\omega_L$ its frequency.
Moving into the frame of the drive, the full Hamiltonian reads
\begin{equation}
\tilde{H} = -\Delta_0 \hat{a}^\dagger \hat{a} + \Omega_m \hat{b}^\dagger \hat{b} - g_0 \hat{a}^\dagger \hat{a} (\hat{b}+\hat{b}^\dagger) + \Omega (\hat{a} + \hat{a}^\dagger)
\end{equation}
where we have introduced the detuning between the drive and the resonator $\Delta_0 = \omega_L - \omega_R$.

The classical nonlinear equations of motion in the frame of the drive, i.e.~Equations (\ref{eq:classicala}) and (\ref{eq:classicalb}) in the main text, are now easily obtained from the Heisenberg equations of motion for the operators, $\dot{\hat{a}} = -i [\hat{a},\tilde{H}]$, by taking expectation values and factorizing averages, e.g.~$\langle \hat{a} \hat{b} \rangle \rightarrow \langle \hat{a} \rangle \langle \hat{b} \rangle$. Including cavity and mechanical damping, at rates $\kappa$ and $\Gamma_m$, we obtain
\begin{eqnarray}
\dot{\bar{a}} & = & +i\Delta_{0}\bar{a}-\frac{\kappa}{2}\bar{a}+ig_{0}\bar{a}(\bar{b}+\bar{b}^{\star}) + \sqrt{\kappa} \bar{a}_\textrm{in} \label{eq:classicalaSI}\\
\dot{\bar{b}} & = & -i\Omega_{m}\bar{b}-\frac{\Gamma_{m}}{2}\bar{b}+ig_{0}|\bar{a}|^{2}
\label{eq:classicalbSI}
\end{eqnarray}
where we introduced $\langle \hat{a} \rangle = \bar{a}$, $\langle \hat{b} \rangle = \bar{b}$, and $-i \Omega = +\sqrt{\kappa} \bar{a}_\textrm{in}$.

Fluctuations around the amplitudes $\bar{a}$ and $\bar{b}$ are described by bosonic operators $\delta\hat{a}$ and $\delta\hat{b}$ obeying linear quantum Langevin equations. They can be derived from $\hat{a} = e^{-i \omega_L t} (\bar{a} + \delta\hat{a})$ and $\hat{b} = \bar{b} + \delta\hat{b}$ where $\bar{a}$ and $\bar{b}$ obey the nonlinear classical equations of motion (\ref{eq:classicala}) and (\ref{eq:classicalb}). Neglecting nonlinear terms we get
\begin{eqnarray}
\dot{\delta\hat{a}} & = & +i\Delta \delta\hat{a} -\frac{\kappa}{2} \delta\hat{a} +iG(\delta\hat{b}+\delta\hat{b}^{\dagger})+\sqrt{\kappa}\,\hat{a}_{\text{in}} \label{eq:langevindSI}\\
\dot{\delta\hat{b}} & = & -i\Omega_{m} \delta\hat{b}-\frac{\Gamma_{m}}{2}\delta\hat{b} +iG(\delta\hat{a}+\delta\hat{a}^{\dagger})+\sqrt{\Gamma_{m}}\,\hat{b}_{\text{in}} \label{eq:langevincSI}
\end{eqnarray}
where $\Delta=\Delta_{0}+g_{0}(\bar{b}+\bar{b}^{\star})$ denotes the shifted detuning and $G=g_{0}|\bar{a}|$ is the enhanced optomechanical coupling.

\section{Details on the exact solution and relation to the quantum noise approach (5)}

Our treatment of the linear Langevin equations (\ref{eq:langevind}) and (\ref{eq:langevinc}) follows closely the methodology and notation of e.g.~Ref.~\cite{Marquardt2007}.
While in the NDR it is most natural to solve the set of coupled equations in terms of a modified \emph{mechanical} response due to an \emph{optical} self-energy, in the RDR we write down a modified \emph{optical} response due to a \emph{mechanical} self-energy.

Equations (\ref{eq:langevind}) and (\ref{eq:langevinc}) can be solved in the Fourier domain
\begin{align}
\delta\hat{b}[\omega] & = \chi_m[\omega] \left[\sqrt{\Gamma_m}\,\hat{b}_\textrm{in}[\omega]
+i G (\delta\hat{a}[\omega]+\delta\hat{a}^\dagger[\omega])\right]\\
\delta\hat{b}^\dagger[\omega] & = \chi_m^\star[-\omega] \left[\sqrt{\Gamma_m}\,\hat{b}^\dagger_\textrm{in}[\omega]
-i G (\delta\hat{a}[\omega]+\delta\hat{a}^\dagger[\omega])\right]
\end{align}
and
\begin{align}
\label{solution}
& \left(
\begin{matrix}
\delta\hat{a}[\omega] \\
\delta\hat{a}^\dagger[\omega]
\end{matrix}
\right) \\
& = \frac{\sqrt{\kappa}}{\mathcal{N}[\omega]}
\left(
\begin{matrix}
\chi^{\star -1}_R[-\omega] - i \Sigma[\omega] & -i\Sigma[\omega]\\
+i\Sigma[\omega] & \chi^{-1}_R[+\omega] + i\Sigma[\omega]
\end{matrix}
\right)
\left(
\begin{matrix}
\delta\hat{a}_\textrm{in}[\omega] \\
\delta\hat{a}^\dagger_\textrm{in}[\omega]
\end{matrix}
\right) \nonumber \\
& + \frac{i G \sqrt{\Gamma_m}}{\mathcal{N}[\omega]}
\left(
\begin{matrix}
\chi_R^{\star -1}[-\omega] \chi_m[\omega] & \chi_R^{\star -1}[-\omega] \chi_m^\star[-\omega]\\
-\chi_R^{-1}[+\omega] \chi_m[\omega] & -\chi_R^{-1}[+\omega] \chi_m^\star[-\omega]
\end{matrix}
\right)
\left(
\begin{matrix}
\delta\hat{b}_\textrm{in}[\omega] \\
\delta\hat{b}^\dagger_\textrm{in}[\omega]
\end{matrix}
\right) \nonumber
\end{align}
with the \emph{mechanical} self-energy
\begin{equation}
\Sigma[\omega]=-iG^{2}(\chi_{m}[\omega]-\chi_{m}^{\star}[-\omega]) = \Sigma^\star[-\omega]
\label{self-energy}
\end{equation}
modifying the \emph{optical} response,
\begin{equation}
\mathcal{N}[\omega]=\chi^{-1}_{R}[\omega]\chi^{\star-1}_{R}[-\omega] -2\Delta \Sigma[\omega] = \mathcal{N}^\star[-\omega],
\end{equation}
the mechanical and the resonator response function, $ \chi_{m}[\omega]=[\Gamma_{m}/2-i(\omega-\Omega_{m})]^{-1}$ and $\chi_{R}[\omega]=[\kappa/2-i(\omega+\Delta)]^{-1}$.

Although this solution has a different form as compared to the one in e.g.~Ref.~\cite{Marquardt2007}, it is the identical analytical expression. In its current form, however, we can readily connect the exact expression for the self-energy (\ref{self-energy}) to the change in the cavity linewidth $\kappa_\textrm{om}$ and the detuning $\Delta_{\textrm{om}}$ we find from the quantum noise approach, i.e.~Eq.~(\ref{eq:kappaeff}) of the main text, as follows
\begin{align}
\kappa_{\textrm{om}} & = -2\textrm{Im}\,\Sigma[-\Delta]\\
& = \frac{\Gamma_{m}G^{2}}{\frac{\Gamma_{m}^{2}}{4}+(\Delta+\Omega_{m})^{2}}-\frac{\Gamma_{m}G^{2}}{\frac{\Gamma_{m}^{2}}{4}+(\Delta-\Omega_{m})^{2}}
\end{align}
with $\kappa_{\textrm{eff}}=\kappa+\kappa_{\textrm{om}}$ and $\Delta_{\textrm{eff}}=\Delta+\Delta_{\textrm{om}}$ with
\begin{align}
\Delta_{\textrm{om}} & =-\textrm{Re}\,\Sigma[-\Delta]\\
& = \frac{\Gamma_{m}G^{2} (\Delta+\Omega_m)}{\frac{\Gamma_{m}^{2}}{4}+(\Delta+\Omega_{m})^{2}}-\frac{\Gamma_{m}G^{2}(\Delta-\Omega_m)}{\frac{\Gamma_{m}^{2}}{4}+(\Delta-\Omega_{m})^{2}}.
\end{align}

\section{Details on input-output solution (6)}

The set of linear quantum Langevin equation (\ref{eq:langevind}) and (\ref{eq:langevinc}) for the fluctuation operators can also be written in matrix form
\begin{equation}
\dot{\mathbf{u}} = \textbf{\texttt{M}} \, \mathbf{u}
+ \textbf{\texttt{L}} \, \mathbf{u}_\textrm{in}
\end{equation}
with $\mathbf{u} = (\delta \hat{a}, \delta \hat{a}^\dagger, \delta \hat{b}, \delta \hat{b}^\dagger)^T$, $\mathbf{u}_\textrm{in} = (\hat{a}_\textrm{in}, \hat{a}^\dagger_\textrm{in}, \hat{b}_\textrm{in}, \hat{b}^\dagger_\textrm{in})^T$ and
\begin{equation}
\textbf{\texttt{M}} =
\left(
\begin{matrix}
i\Delta-\frac{\kappa}{2} & 0 & iG & iG \\
0 & -i\Delta-\frac{\kappa}{2} & -iG & -iG \\
iG & iG & -i\Omega_m-\frac{\Gamma_m}{2} & 0\\
-iG & -iG & 0 & i\Omega_m-\frac{\Gamma_m}{2}
\end{matrix}
\right)
\end{equation}
as well as
$\textbf{\texttt{L}} = \textrm{Diag}[\sqrt{\kappa},\sqrt{\kappa},\sqrt{\Gamma_m},\sqrt{\Gamma_m}]$.

Using the input-output relations for a one-sided cavity
\begin{equation}
\mathbf{u}_{\textrm{out}} = \mathbf{u}_{\textrm{in}}-\textbf{\texttt{L}} \, \mathbf{u}
\label{inputoutput}
\end{equation}
we can solve the input-output problem in the Fourier domain
\begin{equation}
\mathbf{u}_{\textrm{out}}(\omega)=\textbf{\texttt{U}}(\omega)\, \mathbf{u}_{\textrm{in}}(\omega)
\end{equation}
with
\begin{equation}
\textbf{\texttt{U}}(\omega) = \mathbb{1}_{4\times4} + \textbf{\texttt{L}} \, [+i \omega \mathbb{1}_{4\times4} + \textbf{\texttt{M}}]^{-1} \, \textbf{\texttt{L}}.
\end{equation}
Equation (\ref{eq:inoutsolution}) in the main text is the first row of $U(\omega)$, i.e.
\begin{equation}
\hat{a}_{\textrm{out}}=A(\omega)\hat{a}_{\textrm{in}}+B(\omega)\hat{a}_{\textrm{in}}^{\dagger}+C(\omega)\hat{b}_{\textrm{in}}+D(\omega)\hat{b}_{\textrm{in}}^{\dagger}
\end{equation}
with $A(\omega) = \textbf{\texttt{U}}_{11}$, $B(\omega) = \textbf{\texttt{U}}_{12}$, $C(\omega) = \textbf{\texttt{U}}_{13}$, $D(\omega) = \textbf{\texttt{U}}_{14}$.

We can express the coefficients in terms of the solution (\ref{solution})
\begin{align}
A(\omega) &= 1 - \frac{\kappa}{\mathcal{N}[\omega]} \left( \chi^{\star -1}_R[-\omega] - i \Sigma[\omega] \right)\\
B(\omega) &= i\kappa \frac{\Sigma[\omega]}{\mathcal{N}[\omega]}\\
C(\omega) &= \frac{i G \sqrt{\kappa\Gamma_m}}{\mathcal{N}[\omega]}
\chi_R^{\star -1}[-\omega] \chi_m[\omega]\\
D(\omega) &= -\frac{i G \sqrt{\kappa\Gamma_m}}{\mathcal{N}[\omega]}
\chi_R^{\star -1}[-\omega] \chi_m^\star[-\omega].
\end{align}

\section{Details on the three-mode model}

In this section we give the Hamiltonian and the set of quantum Langevin equations for the  three-mode model discussed in the later part of the main text.

As mentioned in the main text it is advantageous to realize the RDR not with a low-$Q$ mechanical oscillator that consequentially has a large mechanical decoherence rate $\Gamma_{m}\bar{n}_{\textrm{th}}$, but with a high-$Q$ mechanical oscillator whose damping is increased by sideband cooling \cite{Schliesser2008}, i.e.~a mechanical oscillator with large quality factor $\Gamma_{m}\ll\kappa$ is coupled to a second electromagnetic mode. This is shown in Fig.~\ref{fig:setup} (b) and (c).

The linearized Hamiltonian for the three-mode system in the rotating frame reads ($\hbar=1$)
\begin{align}
H = & -\Delta \delta\hat{a}_1^\dagger \delta\hat{a}_1 -\Delta_2 \delta\hat{a}_2^\dagger \delta\hat{a}_2 + \Omega_m \hat{b}^\dagger \hat{b} \nonumber \\
&+ G (\hat{a}_1^\dagger \hat{b} + \textrm{H.c.}) + G_2 (\hat{a}_2^\dagger \hat{b} + \textrm{H.c.})
\end{align}
where we have neglected the mechanically-induced coupling of the optical modes \cite{Law1995, Dobrindt2010}.

Analogous to Eqns.~(\ref{eq:langevind}) and (\ref{eq:langevinc}) the equations of motion for the fluctuations of the three-mode model read
\begin{align}
\delta\dot{\hat{a}}_1 = & +i\Delta \delta\hat{a}_1 -\frac{\kappa}{2} \delta\hat{a}_1 +iG(\delta\hat{b}+\delta\hat{b}^{\dagger})+\sqrt{\kappa}\,\hat{a}_{\text{in}}^{(1)}\\
\delta\dot{\hat{a}}_2 = & +i\Delta_2 \delta\hat{a}_2 -\frac{\kappa_2}{2} \delta\hat{a}_2
+iG_2(\delta\hat{b}+\delta\hat{b}^{\dagger})+\sqrt{\kappa_2}\,\hat{a}_{\text{in}}^{(2)}\\
\delta\dot{\hat{b}} = & -i\Omega_{m} \delta\hat{b}
-\frac{\Gamma_{m}}{2}\delta\hat{b} + \sqrt{\Gamma_{m}}\,\hat{b}_{\text{in}} \nonumber \\
& +iG(\delta\hat{a}_1+\delta\hat{a}^{\dagger}_1) +iG_2(\delta\hat{a}_2+\delta\hat{a}_2^{\dagger})
\end{align}
where we have neglected the coupling between the optical modes introduced by the common dissipation channel \cite{Dobrindt2010}.

\begin{table*}
\caption{Parameters of existing and perspective superconducting optomechanical systems.}
\begin{ruledtabular}
\begin{tabular}{ccccccccc}
&${\omega_c}/2\pi$&${\kappa}/2\pi$&${\Omega_m}/2\pi$&
${\Gamma_m}/2\pi$&${g_0}/2\pi$ & $\bar{n}_p$ & $\bar{n}_{p2}$ & $P_{\rm in},\,P_{\rm in\_2}$ (nW)\\
\hline
Teufel et al.~\cite{Teufel2011}&$7.5$ GHz&$170$ kHz&$10$ MHz&$30$ Hz&$230$ Hz&$1.3\cdot10^6$&$2.6\cdot10^7$& 100\\
TiN (proposed) &$7.5$ GHz&$10$ kHz&$1$ MHz&$50$ Hz&$100$ Hz & $2.5\cdot10^4$&$5\cdot10^5$& 0.3
\label{tab:table1}
\end{tabular}
\end{ruledtabular}
\end{table*}

The solution in Fourier space with the input-output relation (\ref{inputoutput}) can be written in matrix form
\begin{equation}
\mathbf{u}_{\textrm{out}}(\omega)=\textbf{\texttt{U}}(\omega)\, \mathbf{u}_{\textrm{in}}(\omega)
\end{equation}
with
\begin{align}
\mathbf{u} = &(\delta \hat{a}_1, \delta \hat{a}^\dagger_1, \delta \hat{a}_2, \delta \hat{a}^\dagger_2, \delta \hat{b}, \delta \hat{b}^\dagger)^T \\
\mathbf{u}_\textrm{in} = &(\hat{a}_\textrm{in}^{(1)}, (\hat{a}_\textrm{in}^{(1)})^\dagger, \hat{a}_\textrm{in}^{(2)}, (\hat{a}_\textrm{in}^{(2)})^\dagger, \hat{b}_\textrm{in}, \hat{b}^\dagger_\textrm{in})^T
\end{align}
\begin{align}
\textbf{\texttt{U}}(\omega) = \mathbb{1}_{6\times6} + \textbf{\texttt{L}} \, [+i \omega \mathbb{1}_{6\times6} + \textbf{\texttt{M}}]^{-1} \, \textbf{\texttt{L}}
\end{align}
with
\begin{widetext}
\begin{equation}
\textbf{\texttt{M}} =
\left(
\begin{matrix}
  i\Delta-\frac{\kappa}{2} & 0 & 0 & 0 & iG & iG \\
  0 & -i\Delta-\frac{\kappa}{2} & 0 & 0& -iG & -iG \\
  0 & 0 & i\Delta_2-\frac{\kappa_2}{2} & 0 & iG_2 & iG_2 \\
  0 & 0 & 0 & -i\Delta_2-\frac{\kappa_2}{2}& -iG_2 & -iG_2 \\
  iG & iG & iG_2 & iG_2 & -i\Omega_m-\frac{\Gamma_m}{2} & 0\\
  -iG & -iG & -iG_2 & -iG_2 & 0 & i\Omega_m-\frac{\Gamma_m}{2}
 \end{matrix}
\right)
\end{equation}
\end{widetext}
and
$\textbf{\texttt{L}} = \textrm{Diag}[\sqrt{\kappa},\sqrt{\kappa},\sqrt{\kappa_2},\sqrt{\kappa_2},\sqrt{\Gamma_m},\sqrt{\Gamma_m}]$.

If the electromagnetic mode $\hat{a}_2$ is in the resolved sideband limit $\Omega_m \gg \kappa_2$, driven on the red sideband $\Delta_2 = -\Omega_m$, and the coupling $G_2$ is weak, the main effect of the optomechanical coupling is to renormalize the mechanical line width
\begin{equation}
\Gamma_\textrm{eff} = \Gamma_m - \frac{G^2}{\kappa} = (1+\mathcal{C}_2)\Gamma_m.
\end{equation}
For $\mathcal{C}_{2}=4G_{2}^{2}/(\kappa_{2}\Gamma_{m})\gg1$, but within the weak coupling regime, i.e.~$G_2 \ll \kappa_2$, the effective mechanical linewidth is
\begin{equation}
\Gamma_{\textrm{eff}}\sim \frac{\kappa_{2}}{2}.
\end{equation}
Cooling therefore establishes the RDR with respect to the electromagnetic mode $\hat{a}_1$ as the mechanical damping satisfies $\Gamma_{\textrm{eff}}\sim\kappa_{2}/2\gg\kappa$. In addition to reaching the RDR, this has the advantage of cooling the mechanical oscillator down to an effective phonon number
\begin{equation}
\bar{n}_{\textrm{eff}}=\frac{\bar{n}_{\textrm{th}}}{\mathcal{C}_{2}+1}+\frac{\mathcal{C}_2}{\mathcal{C}_2+1} \cdot\frac{\kappa_{2}^{2}}{16\Omega_{m}^{2}}.
\end{equation}

\section{Details on experimental feasibility}

In Table~\ref{tab:table1} we give parameters of superconducting optomechanical systems reported in the literature based on Al as well as parameters of a perspective structure based on recent advances in the fabrication of ultra high-$Q$ TiN microwave cavities. This material is highly promising in the current endeavor due to its very high quality factor, corresponding to a cavity decay rate of $ \kappa/{2\pi} \approx\mathcal{O}(100) \mathrm{Hz}$. 

To evaluate if any of these structures can be used to realize the proposed amplifier scheme we assume that each system has a second mode (``cooling'' mode) with a much larger decay rate $\kappa_2 \approx 20\kappa \gg \kappa$.  Such a scenario can be realized by coupling one mechanical oscillator to two superconducting resonators, one of which is strongly over-coupled, while the second electromagnetic mode is weakly-coupled via the circuit shown in Fig.~5 of the main manuscript. Moreover, we assume optomechanical vacuum coupling rates ($g_0$) that have been demonstrated in the literature for vibrating gap capacitors. To gauge the validity of the scheme, we first compute the power necessary to cool the mechanical oscillator via the ''cooling'' mode (with linewidth $\kappa_2$ ) to a regime where its effective mechanical damping rate $\Gamma_{\rm eff}$ exceeds the linewidth of the first electromagnetic mode, i.e.~$\Gamma_{\rm eff} \approx 10 \kappa \gg \kappa$. In the resolved sideband regime ($\kappa_2\gg\Omega_m$), the cooling rate is given by $\Gamma_{\rm eff} \approx \frac{4 g_0^2 \bar{n}_{p2}}{\kappa_2}$ and the required pump power that needs to be launched on the lower sideband of the auxiliary resonance is therefore given by
\begin{equation}
P_{\rm in\_2} = \frac{\hbar \omega_{R2} \Gamma_{\rm eff} \Omega_m^2}{g_0^2}
\end{equation}
where $\omega_{R2}$ is the resonance frequency of the ``cooling'' mode. For simplicity we will assume that $\omega_{R2} \sim \omega_R$. Table~\ref{tab:table1} shows the required pump cooling power $P_{\rm in\_2}$ and intra-cavity photon number $\bar{n}_{p2}$ for cooling to $\Gamma_{\rm eff} \approx 10 \kappa \gg \kappa$. For a superconducting circuit optomechanical system in a He3/He4 dilution refridgerator ($T<20 \rm  mK$), this cooling rate is sufficient to cool the oscillator to the quantum ground-state given that $\Gamma_{\rm eff} = 10 \kappa \gg \Gamma_m \cdot \bar{n}_m$ where $\bar{n}_m \approx \frac{k_B T}{\hbar \Omega_m} = \mathcal{O}(100)$ is the thermal occupancy of the mechanical oscillator. The latter ensures that the amplifier is only limited by quantum noise. 

Note that for the calculated pump powers and photon numbers, the resonators are still expected to exhibit superconducting behavior (i.e.~the power levels are below the critical current density of Al and TiN).

After establishing the reversed dissipation hierarchy for the first electromagnetic mode by opto-mechanical sideband cooling, i.e.~$ \Gamma_{\rm eff} \gg \kappa$, we can next calculate the power necessary to achieve amplification, by considering the necessary power that needs to be launched on the upper sideband of the first mode. The maximum permissible pump power to operate the system as an amplifier is given by the power that is required to reach the parametric instability, which would lead to coherent emission of microwaves (i.e.~the analog of a Brillouin laser in the optomechanical domain). As mentioned in the main text, the critical mean photon number to undergo self-sustained oscillations is given by
\begin{equation}
\frac{4g_0^2\bar{n}_p}{\kappa \Gamma_{\rm eff}}= 1.
\end{equation}
In the resolved sideband regime, the input pump power required to reach this mean photon number is
\begin{equation}
P_{\rm in} = \frac{\hbar \omega_R \Gamma_{\rm eff} \Omega_m^2}{g_0^2}.
\end{equation}
The pump powers on the upper sideband necessary to reach the parametric instability threshold are given in Table~\ref{tab:table1}. One can readily see that these powers are similar to ones required to cool the mechanical oscillator to $\Gamma_{\rm eff} \gg \Gamma_m$ and thus do not destroy superconductivity in the resonator. Therefore, the proposed amplification scheme can be realized using the currently available technology~\cite{Teufel2011}.

\end{document}